\input harvmac

\noblackbox

\def\sech{\mathop{\rm sech}\nolimits}
\def\J{{\cal{J}}}
\def\inbar{\vrule height1.5ex width.4pt depth0pt}
\def\IR{{\relax\,\hbox{$\inbar\kern-.3em{\rm R}$}}}
\def\IP{{\relax\,\hbox{$\inbar\kern-.3em{\rm P}$}}}
\def\Zbar{{\bar{Z}}}
\def\lambdabar{{\bar{\lambda}}}
\def\hapone{\textstyle{p_1 \over 2}}
\def\haptwo{\textstyle{p_2 \over 2}}
\def\p{\partial}
\def\pbar{{\bar{\partial}}}
\def\zbar{{\bar{z}}}
\def\ybar{{\bar{y}}}
\def\D{{\cal{D}}}
\def\Dbar{{\overline{\cal{D}}}}

\lref\HofmanXT{
  D.~M.~Hofman and J.~M.~Maldacena,
  ``Giant magnons,''
  J.\ Phys.\ A {\bf 39}, 13095 (2006)
  [arXiv:hep-th/0604135].
  %%CITATION = HEP-TH 0604135;%%
}

\lref\DoreyDQ{
  N.~Dorey,
  ``Magnon bound states and the AdS/CFT correspondence,''
  J.\ Phys.\ A {\bf 39}, 13119 (2006)
  [arXiv:hep-th/0604175].
  %%CITATION = HEP-TH 0604175;%%
}

\lref\ChenGE{
  H.~Y.~Chen, N.~Dorey and K.~Okamura,
  ``Dyonic giant magnons,''
  JHEP {\bf 0609}, 024 (2006)
  [arXiv:hep-th/0605155].
  %%CITATION = HEP-TH 0605155;%%
}

\lref\ArutyunovGS{
  G.~Arutyunov, S.~Frolov and M.~Zamaklar,
  ``Finite-size effects from giant magnons,''
  arXiv:hep-th/0606126.
  %%CITATION = HEP-TH 0606126;%%
}

\lref\MinahanBD{
  J.~A.~Minahan, A.~Tirziu and A.~A.~Tseytlin,
  ``Infinite spin limit of semiclassical string states,''
  JHEP {\bf 0608}, 049 (2006)
  [arXiv:hep-th/0606145].
  %%CITATION = HEP-TH 0606145;%%
}

\lref\ChuAE{
  C.~S.~Chu, G.~Georgiou and V.~V.~Khoze,
  ``Magnons, classical strings and beta-deformations,''
  arXiv:hep-th/0606220.
  %%CITATION = HEP-TH 0606220;%%
}

\lref\BobevFG{
  N.~P.~Bobev and R.~C.~Rashkov,
  ``Multispin giant magnons,''
  Phys.\ Rev.\ D {\bf 74}, 046011 (2006)
  [arXiv:hep-th/0607018].
  %%CITATION = HEP-TH 0607018;%%
}

\lref\KruczenskiPK{
  M.~Kruczenski, J.~Russo and A.~A.~Tseytlin,
  ``Spiky strings and giant magnons on $S^5$,''
  arXiv:hep-th/0607044.
  %%CITATION = HEP-TH 0607044;%%
}

\lref\BozhilovBI{
  P.~Bozhilov and R.~C.~Rashkov,
  ``Magnon-like dispersion relation from M-theory,''
  arXiv:hep-th/0607116.
  %%CITATION = HEP-TH 0607116;%%
}

\lref\HuangVZ{
  W.~H.~Huang,
  ``Giant magnons under NS-NS and Melvin fields,''
  arXiv:hep-th/0607161.
  %%CITATION = HEP-TH 0607161;%%
}

\lref\GomezVA{
  C.~Gomez and R.~Hernandez,
  ``The magnon kinematics of the AdS/CFT correspondence,''
  arXiv:hep-th/0608029.
  %%CITATION = HEP-TH 0608029;%%
}

\lref\ChenGQ{
  H.~Y.~Chen, N.~Dorey and K.~Okamura,
  ``On the scattering of magnon boundstates,''
  arXiv:hep-th/0608047.
  %%CITATION = HEP-TH 0608047;%%
}

\lref\RoibanGS{
  R.~Roiban,
  ``Magnon bound-state scattering in gauge and string theory,''
  arXiv:hep-th/0608049.
  %%CITATION = HEP-TH 0608049;%%
}

\lref\OkamuraZV{
  K.~Okamura and R.~Suzuki,
  ``A perspective on classical strings from complex sine-Gordon solitons,''
  arXiv:hep-th/0609026.
  %%CITATION = HEP-TH 0609026;%%
}

\lref\HiranoTI{
  S.~Hirano,
  ``Fat magnon,''
  arXiv:hep-th/0610027.
  %%CITATION = HEP-TH 0610027;%%
}

\lref\RyangYQ{
  S.~Ryang,
  ``Three-spin giant magnons in $AdS_5 \times S^5$,''
  arXiv:hep-th/0610037.
  %%CITATION = HEP-TH 0610037;%%
}

\lref\ChenGP{
  H.~Y.~Chen, N.~Dorey and K.~Okamura,
  ``The Asymptotic Spectrum of the N=4 Super Yang-Mills Spin Chain,''
  arXiv:hep-th/0610295.
  %%CITATION = HEP-TH 0610295;%%
}

\lref\SpradlinWK{
  M.~Spradlin and A.~Volovich,
  ``Dressing the giant magnon,''
  JHEP {\bf 0610}, 012 (2006)
  [arXiv:hep-th/0607009].
  %%CITATION = HEP-TH 0607009;%%
}

\lref\PohlmeyerNB{
  K.~Pohlmeyer,
  ``Integrable Hamiltonian Systems And Interactions Through Quadratic
  Constraints,''
  Commun.\ Math.\ Phys.\  {\bf 46}, 207 (1976).
  %%CITATION = CMPHA,46,207;%%
}

\lref\MikhailovZD{
  A.~Mikhailov,
  ``B\"acklund transformations, energy shift and the plane wave limit,''
  arXiv:hep-th/0507261.
  %%CITATION = HEP-TH 0507261;%%
}

\lref\ZakharovPP{
  V.~E.~Zakharov and A.~V.~Mikhailov,
  ``Relativistically Invariant Two-Dimensional Models In Field Theory
  Integrable By The Inverse Problem Technique. (In Russian),''
  Sov.\ Phys.\ JETP {\bf 47}, 1017 (1978)
  [Zh.\ Eksp.\ Teor.\ Fiz.\  {\bf 74}, 1953 (1978)].
  %%CITATION = SPHJA,47,1017;%%
}

\lref\ZakharovTY{
  V.~E.~Zakharov and A.~V.~Mikhailov,
  ``On The Integrability Of Classical Spinor Models In Two-Dimensional
  Space-Time,''
  Commun.\ Math.\ Phys.\  {\bf 74}, 21 (1980).
  %%CITATION = CMPHA,74,21;%%
}

\lref\HarnadWE{
  J.~P.~Harnad, Y.~Saint Aubin and S.~Shnider,
  ``Backlund Transformations For Nonlinear Sigma Models With Values In
  Riemannian Symmetric Spaces,''
  Commun.\ Math.\ Phys.\  {\bf 92}, 329 (1984).
  %%CITATION = CMPHA,92,329;%%
}

\lref\BeisertTM{
  N.~Beisert,
  ``The su$(2|2)$ dynamic S-matrix,''
  arXiv:hep-th/0511082.
  %%CITATION = HEP-TH 0511082;%%
}

\lref\JanikDC{
  R.~A.~Janik,
  ``The $AdS_5 \times S^5$
  superstring worldsheet S-matrix and crossing symmetry,''
  Phys.\ Rev.\ D {\bf 73}, 086006 (2006)
  [arXiv:hep-th/0603038].
  %%CITATION = HEP-TH 0603038;%%
}

\Title
{\vbox{
\baselineskip12pt
\hbox{hep-th/0611033}
}}
{\vbox{
\centerline{Dressing the Giant Magnon II}
}}

\centerline{
Chrysostomos Kalousios,
Marcus Spradlin
and Anastasia Volovich
}

\bigskip

\centerline{
{\tt chrysostomos\_kalousios@brown.edu},
{\tt spradlin@het.brown.edu},
{\tt nastja@het.brown.edu}
}

\vskip .5in
\bigskip
\centerline{Brown University}
\centerline{Providence, Rhode Island 02912 USA}

\vskip .5in
\centerline{\bf Abstract}

We extend our earlier work 
by demonstrating how to construct classical string solutions
describing arbitrary superpositions of scattering and bound states
of dyonic giant magnons on $S^5$ using the dressing method for
the $SU(4)/Sp(2)$ coset model.
We present a particular scattering solution
which generalizes solutions
found in hep-th/0607009 and hep-th/0607044
to the case of arbitrary magnon momenta.
We compute the classical time delay for the scattering
of two dyonic magnons carrying angular momenta with arbitrary
relative orientation on the $S^5$.

\Date{}

\listtoc
\writetoc

\newsec{Introduction}

The study of classical spinning string solutions
in $AdS_5 \times S^5$ has provided a wealth of data for
detailed study of the AdS/CFT correspondence.
An interesting step forward was recently taken by Hofman
and Maldacena \HofmanXT, who found the classical string solution
corresponding
to a single magnon in the dual gauge theory.
In this context the word magnon refers to an elementary excitation
which can travel along a chain of $Z$'s with some momentum $p$,
i.e.
\eqn\aaa{
{\cal O}_p \sim \sum_l e^{i p l} ( \cdots Z Z Z W Z Z Z \cdots ),
}
where the magnon $W$ is
inserted at position $l$ along the chain.
The corresponding `giant magnon' is an open string whose endpoints
move at the speed of light along an equator of the $S^5$, separated
in longitude by an angle $p$. 
This state carries an infinite amount of angular momentum $J$ in the plane of
the equator of the $S^5$ and is characterized by a finite value of
$\Delta - J$. Recent papers on giant magnons include 
\refs{\DoreyDQ,\ChenGE,\ArutyunovGS,\MinahanBD,\ChuAE,
\BobevFG,\SpradlinWK,\KruczenskiPK,\BozhilovBI,\HuangVZ,\GomezVA,
\ChenGQ,\RoibanGS,\OkamuraZV,\HiranoTI,\RyangYQ,\ChenGP}.

In \SpradlinWK\ we
employed the dressing method
\refs{\ZakharovPP,\ZakharovTY,
\HarnadWE}
to construct classical string solutions corresponding to
various scattering and bound states of magnons.
In particular, we demonstrated how to obtain solutions
representing superpositions of any number of elementary
giant magnons (or bound states thereof)
on $\IR \times S^5$, as well as any number of
dyonic giant magnons on $\IR \times S^3$.
The dyonic giant magnon, discovered in \refs{\DoreyDQ,\ChenGE},
is a BPS bound state of many (${\cal O}(\sqrt{\lambda})$) magnons
which carries,
in addition to an infinite amount of $J$ in the equator of the $S^3$,
a non-zero macroscopic amount of angular momentum $J_1$
in the orthogonal plane on $S^3$.

In this note we study scattering states of dyonic giant magnons
on $S^5$, some special cases of which have appeared
in \refs{\SpradlinWK,\KruczenskiPK,\RyangYQ}.
We fill a gap in our previous work
by demonstrating how to construct classical
string solutions describing general scattering states of dyonic giant
magnons whose individual angular momenta $J_i$ have arbitrary
orientations in the directions transverse to the equator of the $S^5$.

After reviewing the basics of giant magnons in section 2, we explain
in section 3 how to apply the dressing method for the $SU(4)/Sp(2)
= S^5$ coset model, following the construction of \HarnadWE.
This coset construction
apparently has more flexibility than the $SO(6)/SO(5) = S^5$ coset
construction employed in \SpradlinWK, since we have been unable to
find the dyonic giant magnon solution via the latter dressing method.
In \SpradlinWK\ the $SU(2)$ principal
chiral model was instead used
to construct superpositions of dyonic magnons.
That was sufficient for solutions living only on an $S^3 \subset S^5$,
but the $SU(4)/Sp(2)$ coset used in this paper allows us to construct
solutions living on
the full $S^5$.
In section 4 we begin with a detailed analysis of the parameter space
for a single soliton in the $SU(4)/Sp(2)$ coset model.  We present
in (4.13) a particular explicit solution for the scattering
of two dyonic giant magnons
with arbitrary momenta $p_1,p_2$ which carry angular momentum in
orthogonal planes.
This solution generalizes the special case $p_1 = - p_2 = \pi$
which was obtained in \SpradlinWK\ and was generalized to $p_1 = - p_2
=p$ in \KruczenskiPK.
Finally in section 5 we calculate the classical time delay for
the scattering of two dyonic giant magnons with arbitrary relative
orientations on the $S^5$.
It would be interesting to compare the corresponding classical
phase shift to a gauge theory analysis along the lines of
\refs{\HofmanXT,\ChenGQ,\RoibanGS}.

\newsec{Giant Magnons}

We consider string theory on $\IR \times S^5$ in conformal gauge,
writing the $S^5$ part of the theory in terms of three complex fields
$Z_i$ subject to the constraint
\eqn\sphere{
Z_i \Zbar_i = 1.
}
The equation of motion for $Z_i$ can be written as
\eqn\stringeom{
\pbar \p Z_i + \ha (\p Z_j \pbar \Zbar_j+ \p \Zbar_j \pbar Z_j) Z_i = 0,
}
where we use the worldsheet coordinates
$z = \ha (x-t)$, $\zbar = \ha (x+t)$.
The Virasoro constraints take the form
\eqn\virasoro{
\p Z_i \p \bar{Z}_i = \pbar Z_i  \pbar \bar{Z}_i = 1
}
after setting the gauge $X^0 = t$ ($X^0$ is the time coordinate
on $\IR \times S^5$).

We consider a giant magnon to be any open string whose endpoints
move at the speed of light along an equator of the $S^5$, which
we choose lie in the $Z_1$ plane.  The appropriate boundary conditions
at fixed $t$ are
\eqn\magnondef{
\eqalign{
Z_1(t, x \to \pm \infty) &= e^{i( t \pm  p/2) +i  \alpha}, \cr
Z_i(t, x \to \pm \infty) &= 0, \qquad i = 2,3,
}}
where $e^{i\alpha}$ is an arbitrary
overall phase and $p$ represents the difference
in longitude between the endpoints of the string on the equator of the
$S^5$.  In the gauge theory picture, $p$ is identified with the
momentum of the magnon \HofmanXT.
We may refer to $p$ as the `momentum' of a magnon, but it should be
kept in mind that the worldsheet momentum of all of the solutions
we consider is zero due to the Virasoro 
constraints \virasoro.

The equations \sphere--\magnondef\ have
infinitely many distinct solutions, which can be partly classified
by their conserved charges.
The boundary conditions \magnondef\ explicitly break the
$SO(6)$ symmetry of the $S^5$ down to $U(1)\times SO(4)$.
The conserved charge associated with the $U(1)$ is
\eqn\chargeone{
\Delta - J = {\sqrt{\lambda} \over 2 \pi}
\int_{-\infty}^{+\infty} dx
\, \left( 1 - {\rm Im}[\Zbar_1 \p_t Z_1] \right),
}
where $\sqrt{\lambda}/2 \pi$ is the string tension expressed
in terms of the 't Hooft coupling $\lambda$ of the dual gauge theory.
The $SO(4)$ symmetry leads to conserved angular momentum matrix
\eqn\jmatrix{
J_{ab} = i{\sqrt{\lambda} \over 2 \pi} 
\int_{-\infty}^\infty dx\, \left( X_a \p_t X_b - X_b \p_t X_a\right),
\qquad a,b=1,\ldots,4,
}
which we have written in terms of the real basis defined by
$Z_2 = X_1 + i X_2$, $Z_3 = X_3 + i X_4$.

\newsec{Dressing Method for $S^5=SU(4)/Sp(2)$}

In order to construct solutions of \sphere--\magnondef\ we will
apply the dressing method of
Zakharov and Mikhailov \refs{\ZakharovPP,\ZakharovTY}
for building soliton solutions of classically
integrable equations, following
the application of this method
to the $SU(4)/Sp(2)$ coset model given in \HarnadWE.

We will see that
an elementary soliton of the $SU(4)/Sp(2)$ coset model is
characterized by the choice of a complex parameter $\lambda$ and a point
$w$ on $\IP^3$.
The most general solution obtainable\foot{It is not clear
to us that all solutions may be obtained through the dressing
method.  For example,
we have been unable to obtain the dyonic giant magnon
solution via the dressing method in the $SO(6)/SO(5)$ coset model.}
via the dressing method takes the form of a scattering state of
any number of elementary solitons or bound states of them.
Each individual soliton carries
some `momentum' $p_i$ and a single non-zero $SO(4)$ angular
momentum $J_i$ (i.e., the eigenvalues of the matrix \jmatrix\ for
a single soliton are
$\{ + J_i, - J_i, 0, 0 \}$).  These two quantities are encoded in
the parameter $\lambda_i$ of the soliton, while the parameter
$w_i$ determines
the plane of its angular momentum
in the transverse $\IR^4$ (i.e., the eigenvetors
of \jmatrix).

The simplest context in which the dressing method may be applied
is the
reduced \PohlmeyerNB\ principal chiral model describing
a unitary matrix $g(z,\zbar)$
satisfying the equation of motion
\eqn\eom{
\pbar (\p g \, g^{-1}) + \p (\pbar g \, g^{-1}) = 0
}
subject to the Virasoro constraints
\eqn\virasoro{
(i g^{-1} \p g)^2 = 1, \qquad
(i g^{-1} \pbar g)^2 = 1.
}
Given any solution $g(z,\zbar)$ of these equations, the dressing method
provides for the construction
of an appropriate dressing matrix $\chi$ such that
\eqn\aaa{
g'(z,\zbar) = \chi(z,\zbar) g(z,\zbar)
}
is also solution of \eom\ and \virasoro.

For the application to classical string theory on $\IR \times S^5$
we are not interested in a principal chiral model but rather a coset
model.  In previous work \SpradlinWK\ we employed the $S^5 = SO(6)/SO(5)$
but for the present analysis it is more fruitful to
use the coset $S^5 = SU(4)/Sp(2)$ following the analysis of \HarnadWE.
We define this coset by imposing on $g \in SU(4)$ the constraint
\eqn\cosetcondition{
g^{\rm T} = \J g \J^{-1},
}
where $\J$ is the fixed antisymmetric matrix
\eqn\aaa{
\J = \pmatrix{0 & 0 & +1 & 0 \cr
0 & 0 & 0 & +1 \cr
-1 & 0 & 0 & 0 \cr
0 & -1 & 0 & 0}.
}
A convenient parametrization of this coset, which allows us
to immediately read off the $S^5$ coordinates $Z_i$ from the matrix $g$,
is given by
\eqn\parametrization{
g = \pmatrix{~~Z_1 & ~~Z_2 & 0 & ~~Z_3 \cr
- \Zbar_2 & ~~\Zbar_1 & - Z_3 & 0 \cr
0 & ~~\Zbar_3 & ~~Z_1 & -\Zbar_2 \cr
- \Zbar_3 & 0 & ~~Z_2 & ~~\Zbar_1},
}
which is unitary and satisfies
\cosetcondition\ precisely when \sphere\ holds.

To apply the dressing method, we begin with a given solution
$g$ by solving the linear system
\eqn\aux{
\p \Psi = {\p g \,g^{-1}\Psi\over 1 - \lambda}, \qquad
\pbar \Psi = {\pbar g\,g^{-1}\Psi \over 1 + \lambda}
}
to find $\Psi(\lambda)$ as a function of the auxiliary
complex parameter
$\lambda$, subject to the initial condition
\eqn\initial{
\Psi(0) = g,
}
the unitarity constraint
\eqn\unitarityconstraint{
\left[ \Psi(\lambdabar) \right]^\dagger \Psi(\lambda) = 1,
}
and the coset constraint
\eqn\cosetconstraint{
\Psi(\lambda) = \Psi(0) \J \overline{\Psi(1/\lambdabar)} \J^{-1},
}
whose role is to ensure that the dressed solution $g'$ we now construct
will continue to satisfy the coset condition \cosetcondition.

Once we know $\Psi(\lambda)$, the dressing factor
for a single soliton may be written in terms of the parameters
$(\lambda_i,w_i)$ discussed above as
\HarnadWE
\eqn\aaa{
\chi(\lambda) = 1 + { \lambda_1 - \lambdabar_1 \over \lambda - \lambda_1} P
+ {1/\lambdabar_1 - 1/\lambda_1 \over \lambda - 1/\lambdabar_1} Q,
}
where $P$ is the hermitian projection operator whose image is
spanned by $\Psi(\lambdabar_1) w_1$ for any
constant four-component complex vector $w_1$ (the overall scale
of $w_1$ clearly drops out so it parametrizes $\IP^3$)
and $Q$ is the hermitian projection
operator whose image is spanned by $\Psi(1/\lambda_1) \J \bar{w}_1$.
Concretely,
\eqn\aaa{
P = {
\Psi(\lambdabar_1) w_1 w_1^\dagger \left[ \Psi(\lambdabar_1)\right]^\dagger\over
w_1^\dagger \left[\Psi(\lambdabar_1)\right]^\dagger
\Psi(\lambdabar_1) w_1}, \qquad
Q = { \Psi(1/\lambda_1) \J \bar{w}_1 w_1^{\rm T} \J^{-1}
\left[ \Psi(1/\lambda_1)\right]^\dagger \over
w_1^{\rm T}
\J^{-1} \left[\Psi(1/\lambda_1)\right]^\dagger \Psi(1/\lambda_1) \J \bar{w}_1}.
}
Then
\eqn\aaa{
\Psi'(\lambda) = \chi(\lambda) \Psi(\lambda)
}
satisfies
the constraints \unitarityconstraint\ and \cosetconstraint,
and provides the desired one-soliton solution $g' = \Psi'(0)$ to the original
equations \eom\ and \virasoro.
Unlike the $SO(6)/SO(5)$ coset considered in \SpradlinWK, in this
case there are no restrictions on the complex polarization vector $w_1$.
Repeated application of this procedure
can be used to generate multi-soliton solutions.

\newsec{
Giant Magnons on $\IR \times S^5$}

To apply the dressing method we begin with the vacuum solution
\eqn\vac{
\eqalign{
Z_1 &= e^{i t}, \cr
Z_2 &= 0,\cr
Z_3 &= 0,
}
}
which describes a point-like string moving at the speed of light
around the equator of the $S^5$.
This state clearly has $\Delta - J = 0$.
After embedding this solution into $SU(4)$ as in \parametrization,
a simple calculation reveals that the desired solution $\Psi(\lambda)$
to the linear system
\aux\ subject to the constraints \initial--\cosetconstraint\ is
\eqn\psidef{
\Psi(\lambda) = {\rm diag}(e^{+ i Z(\lambda)},
e^{-i Z(\lambda)}, e^{+i Z(\lambda)}, e^{-i Z(\lambda)}),
\qquad
Z(\lambda) = {z \over \lambda - 1} + {\zbar \over \lambda + 1}.
}

\subsec{A single dyonic giant magnon}

Let us begin by applying the dressing method once
to the vacuum \vac.  We will reproduce the dyonic giant magnon
solution of \ChenGE.
The value of this exercise is to
set some notation for subsequent solutions and also to
illustrate the physical significance of the parameters
$\lambda_1$ and $w_1$ which characterize the soliton.
We parametrize the latter as
\eqn\pthree{
w_1 = \pmatrix{
~+i e^{+ y_1/2 + i \psi_1/2 + i \chi_1/2} \cos \alpha_1 \cr
~~~\,\, e^{- y_1/2 - i \psi_1/2 + i \chi_1/2} \cos \beta_1 \cr
\,-i e^{+ y_1/2 - i \psi_1/2 - i \chi_1/2} \sin \alpha_1 \cr
~~~\,\,e^{- y_1/2 + i \psi_1/2 - i \chi_1/2} \sin \beta_1
},
}
where $y_1$ is complex and the remaining four angles are real.
Here we have used the fact that the overall scale of $w_1$ drops out.
Application of the dressing method gives the solution
\eqn\onesoliton{\eqalign{
Z_1 &= {e^{+i t}  \over |\lambda_1|}
\left[ { \lambdabar_1 e^{-2 i Z(\lambda_1) + \ybar_1} \over \D_1}
+ {\lambda_1 e^{+ 2 i Z(\lambda_1) - \ybar_1} \over \Dbar_1}
\right],\cr
Z_2 &= {i e^{i \psi_1} (\lambdabar_1 - \lambda_1) \over |\lambda_1|}
\left[
{  e^{-i t} \cos \alpha_1 \cos \beta_1 \over \D_1}
+ { e^{+i t} \sin \alpha_1 \sin \beta_1 \over \Dbar_1}
\right],
\cr
Z_3 &= {i e^{i \chi_1} (\lambdabar_1 - \lambda_1) \over |\lambda_1|}
\left[
{e^{-i t} \cos\alpha_1\sin\beta_1 \over \D_1}
- {e^{+i t} \sin\alpha_1 \cos \beta_1 \over \Dbar_1}
\right],
}}
where
\eqn\aaa{
\D_1 = e^{-2 i Z(\lambda_1) + \ybar_1} +
e^{-2 i Z(\lambdabar_1) - y_1}.
}

The solution \onesoliton\ carries $U(1)$ charge
\eqn\dj{
\Delta - J = {\sqrt{\lambda} \over 4 \pi} \left|
\lambda_1 - \lambdabar_1 -
{1 \over \lambda_1} + {1 \over \lambdabar_1}
\right|
}
and one non-zero $SO(4)$ angular momentum
\eqn\jone{
J_1 = {\sqrt{\lambda} \over 4 \pi} \left|
\lambda_1 - \lambdabar_1 +
{1 \over \lambda_1} - {1 \over \lambdabar_1}
\right|.
}
Note that $\Delta - J$ is always strictly positive, but we have
defined $J_1$ to be positive by choice--the eigenvalues of \jmatrix\ come
in $\pm$ pairs.
Furthermore, the value of $p$ for this solution, which may be
read off by comparing \onesoliton\ to \magnondef, is given by
\eqn\pvalue{
e^{i p} = { \lambda_1 \over \lambdabar_1}.
}
In fact, $\lambda_1$ and $\lambdabar_1$ are 
(sometimes up to an author-dependent normalization factor)
the quantities frequently referred to in the recent
literature as $x^+$ and $x^-$ (see in particular
\refs{\BeisertTM,\JanikDC}).
{}From the worldsheet point of view, the solution \onesoliton\ describes
a wave which propagates 
with phase velocity
(i.e., the waveform depends on $x - v_1 t$) given by
\eqn\velocity{
v_1 = {\lambda_1 + \lambdabar_1 \over 1 + |\lambda_1|^2}.
}

Using \dj, \jone\ and \pvalue, we find that the dispersion
relation takes the familiar form
for the dyonic giant magnon \refs{\DoreyDQ,\ChenGE}
\eqn\dispone{
\Delta - J = \sqrt{J_1^2 + {\lambda \over  \pi^2}
\sin^2 {p \over 2}}.
}
Note that all of the parameters associated with the choice of the
`polarization' $w_1$ completely drop out of the expressions for the
conserved charges and the dispersion relation.

We can be more explicit about the role of these parameters.
The parameters $\alpha_1$, $\beta_1$, $\psi_1$ and $\chi_1$
determine the `orientation' of the soliton in the transverse $\IR^4$.
Specifically, the angular momentum matrix \jmatrix\ has
eigenvalues $(+J_1,-J_1,0,0)$, so the soliton is characterized
by an amount $J_1$ of angular momentum inside a certain 2-plane
in the transverse $\IR^4$.  The four parameters $\alpha_1$, $\beta_1$,
$\psi_1$ and $\chi_1$ label the particular plane (they are coordinates
on the Grassmannian ${\rm Gr}_2(\IR^4)$).

The remaining complex parameter $y_1$ can be completely absorbed
by making the translation
\eqn\shift{
Z(\lambda_1) \to Z(\lambda_1) - {i \over 2} \ybar_1.
}
The real part of $y_1$ corresponds to a translation of the soliton
in the $x$ direction while the imaginary part of $y_1$ corresponds
to a rotation inside the plane of the soliton's angular momentum.

\subsec{A scattering state of two dyonic magnons, with three spins on $S^5$}

Having analyzed in detail the parameter space for a single soliton
in the last section, we are now in a position to use the dressing
method to construct multi-soliton scattering states.
The general $n$-soliton solution is specified by $n$ complex
numbers $\lambda_i$ which encode the energy \dj\ and angular momentum
\jone\ of each soliton.  
Each soliton with non-zero angular momentum is also
characterized by the choice of a 2-plane inside the transverse $\IR^4$.
Finally, the $n$-soliton solution has a non-obvious classical
shift symmetry of the form \shift\ for each $i$.  For $n$ solitons
this gives an additional
$2n$ real moduli, but 2 linear combinations can be absorbed
into overall $t$ and $x$ translations.

The procedure for constructing an $n$-soliton solution is therefore
clear, but generic solutions are rather messy.
We display here an explicit formula only for the special case
of two dyonic giant magnons with completely orthogonal
angular momenta, specifically, with soliton 1's angular momentum in
the $Z_2$ plane and soliton 2's angular momentum in the $Z_3$ plane.
To this end we pick the polarization vectors
\eqn\polarizations{
w^{\rm T}_1 = \pmatrix{ i & 1 & 0 & 0 }, \qquad
w^{\rm T}_2 = \pmatrix{ i & 0 & 0 & 1 }.
}
Applying the dressing method twice with parameters
$(\lambda_1,w_1)$ and then $(\lambda_2,w_2)$ gives
the solution
\eqn\twoscattering{\eqalign{
Z_1 &= { e^{+ i t} \over 
|\lambda_1 \lambda_2|}
{{\cal{N}}_{12} \over \D_{12}},
\cr
Z_2 &= {i e^{- i t} (\lambdabar_1 - \lambda_1) \lambda_2 \over
|\lambda_1 \lambda_2|}
\left[
{ \lambda_1 \lambdabar_2 - 1 \over \lambda_1 \lambda_2 - 1}
e^{2 i Z(\lambda_2)}
+  {\lambdabar_1 - \lambdabar_2 \over \lambdabar_1 -
\lambda_2}
e^{2 i Z(\lambdabar_2)}
\right]{ e^{2 i (Z(\lambda_1) + Z(\lambdabar_1))}\over \D_{12}},\cr
Z_3 &= {i e^{- i t} (\lambdabar_2 - \lambda_2) \lambda_1 \over
|\lambda_1 \lambda_2|}
\left[
{ \lambdabar_1 \lambda_2 - 1 \over \lambda_1 \lambda_2 - 1}
e^{2 i Z(\lambda_1)} +
{ \lambdabar_1 - \lambdabar_2 \over \lambda_1 - \lambdabar_2}
e^{2 i Z(\lambdabar_1)}
\right] {e^{2 i (Z(\lambda_2) + Z(\lambdabar_2))}\over \D_{12}},
}}
where
\eqn\aaa{\eqalign{
{\cal{N}}_{12}
= \pmatrix{ \lambda_1 e^{2 i Z(\lambda_1)} &
\lambdabar_1 e^{2 i Z(\lambdabar_1)}}
&\pmatrix{
\left| {\lambda_1 \lambdabar_2 - 1 \over \lambda_1 \lambda_2 - 1}\right|^2
& 1 \cr
1 & \left| {\lambda_1 - \lambda_2 \over \lambda_1 - \lambdabar_2}\right|^2
}
\pmatrix{ \lambda_2 e^{2 i Z(\lambda_2)} \cr
\lambdabar_2 e^{2 i Z(\lambdabar_2)}},\cr
\D_{12} = \pmatrix{ e^{2 i Z(\lambda_1)} & e^{2 i Z(\lambdabar_1)}}
&\pmatrix{
\left| {\lambda_1 \lambdabar_2 - 1 \over \lambda_1 \lambda_2 - 1}\right|^2
& 1 \cr
1 & \left| {\lambda_1 - \lambda_2 \over \lambda_1 - \lambdabar_2}\right|^2
}
\pmatrix{ e^{2 i Z(\lambda_2)} \cr e^{2 i Z(\lambdabar_2)}},
}}
and $Z(\lambda)$ is defined in \psidef.

This solution carries $U(1)$ charge
\eqn\aaa{
\Delta - J = {\sqrt{\lambda} \over 4 \pi} \sum_{i=1}^2
\left| \lambda_i - \lambdabar_i - {1 \over \lambda_i} +
{1 \over \lambdabar_i} \right|
}
and two independent angular momenta
\eqn\aaa{
J_i = {\sqrt{\lambda} \over 4 \pi}
\left| \lambda_i - \lambdabar_i + {1 \over \lambda_i}
- {1 \over \lambdabar_i} \right|,
}
which are the eigenvalues of the angular momentum
matrix \jmatrix\ in the $Z_2$ and $Z_3$ planes respectively.
The total momentum of this giant magnon is
\eqn\aaa{
e^{i p} = 
e^{i (p_1 + p_2)} = {\lambda_1 \over \lambdabar_1} {\lambda_2
\over \lambdabar_2},
}
and the dispersion relation can be written as
\eqn\aaa{
\Delta - J = \sqrt{J_1^2 + {\lambda \over \pi^2}
\sin^2{p_1 \over 2}} +
\sqrt{J_2^2 + {\lambda \over \pi^2} \sin^2{p_2 \over 2}}.
}

Special cases of the solution \twoscattering\ have appeared
previously in the literature.  The case $p_1 = - p_2 = \pi$
was presented in \SpradlinWK, and a generalization to $p_1 = - p_2$
was given in~\KruczenskiPK.
Making direct contact with equation (5.19) of the former 
requires taking the shift parameters $c_i$ in that paper to be
\eqn\aaa{
\tanh c_1 = - \tanh c_2 = - \left|{\lambda_2 \over
\lambda_1}\right| {|\lambda_1|^2 - 1 \over |\lambda_2|^2 - 1}.
}

\subsec{A scattering state of two HM magnons, with arbitrary positions
on the transverse $S^3$}

In the previous subsection we chose
the particular polarization vectors \polarizations\ in order to avoid
too much clutter in \twoscattering.
An interesting limit in which the formulas
simplify is when $|\lambda_i| \to 1$.
Taking $\lambda_i$ onto the unit circle sets the angular momentum
of each soliton to zero--the dyonic giant magnon reduces to the
elementary Hofman-Maldacena magnon~\HofmanXT.
Each such magnon is characterized by a momentum $p$ and a unit
vector $n^a$ in the transverse $\IR^4$ which specifies the polarization
of its fluctuation away from the equator of the $S^5$.
A giant magnon with polarization $n^a$ describes a scalar
field impurity $\phi^a$, $a=1,2,3,4$
in the dual gauge theory.
Using the real basis defined under \jmatrix, the solution for
such a scattering state can be written as
\eqn\hmsolution{\eqalign{
Z_1 &= e^{it} + {e^{it} \over \D_{12}}
\left[ \cos \hapone - \cos \haptwo + i \sin \hapone \tanh u_1
- i \sin \haptwo \tanh u_2\right],
\cr
X^a &= {1 \over \D_{12}} \left[ n_1^a \sin \hapone\sech u_1-
n_2^a \sin \haptwo\sech u_2\right],\qquad  a=1,2,3,4,
}}
where
\eqn\aaa{
u_i = i (Z(\lambda_i) - Z(\lambdabar_i))
= (x - t \cos{\textstyle{p_i \over 2}})\csc {\textstyle{p_i \over 2}}
}
and now
\eqn\aaa{
\D_{12} = {1 - \cos \hapone \cos \haptwo - \sin \hapone \sin \haptwo
\left[\tanh u_1 \tanh u_2 + (n_1 \cdot n_2)
\sech u_1\sech u_2 \right]
\over \cos \haptwo-\cos \hapone}.
}
It is also straightforward to obtain this solution via
the B\"acklund transformation (see \MikhailovZD\ in particular).
The conserved charges and dispersion relation of this solution
do not depend on the polarization vectors $n_i^a$.

\newsec{Classical Time Delay for Scattering of Dyonic Magnons}

With explicit formulas for the scattering solutions in hand, it is
a simple matter to read off the classical time delay for soliton scattering.
To find the time delay as soliton 1 passes soliton 2 (let us take
$v_1 > v_2 > 0$
with the velocities $v_i$ given by \velocity)
we set $x = v_1 (t - \delta t)$ and compare the solution at
$t \to \pm \infty$ to the single soliton solution.
The total time delay is then $\Delta T_{12} = \delta t_+ - \delta t_-$.

We are particularly interested in seeing the dependence of the time
delay on the relative orientations of the angular momenta of the
two scattering solitons.  Without loss of generality we can
take the polarization of $w_1$ as in
\polarizations, but we keep $w_2$ arbitrary as in \pthree.
We find
\eqn\timedelay{
\Delta T_{12} = 
{i \over 2}
{ |1 - \lambda_1|^2 |1 + \lambda_1|^2 \over \lambda_1^2 - \lambdabar_1^2}
\log \left[
(A \cos^2 \alpha_2 + 
 B \sin^2 \alpha_2)
(A \cos^2 \beta_2 + B \sin^2 \beta_2)
\right],
}
where
\eqn\aaa{
A ={|\lambda_1 - \lambda_2|^2
\over |\lambda_1 - \lambdabar_2|^2}, \qquad
B = {|\lambda_1 - 1/\lambdabar_2|^2 \over
|\lambda_1-1/\lambda_2|^2}.
}
It would be interesting to evaluate the corresponding classical
phase shift $\delta_{12}$ (i.e., the $S$-matrix element
$e^{i \delta_{12}}$),
which may be obtained by integrating $\Delta T_{12}$ with respect to
the energy of soliton 1 \dj\ while holding the angular momentum
\jone\ fixed, and to compare the result with a corresponding gauge
theory calculation along the lines of \refs{\ChenGQ,\RoibanGS}.

We can subject \timedelay\ to some consistency checks by comparing
special cases of the formula to known results.
First of all,
we can recover the scattering of two HM magnons by
taking $\lambda_i = e^{i p_i/2}$ on the unit circle.
In this case we obtain
\eqn\hmdt{
\Delta T_{12} = \tan {p_1 \over 2} \log\left[
{1 - \cos{\half (p_1 - p_2)}  \over
1 - \cos{\half (p_1 + p_2)} } 
\right],
}
in complete agreement with the result of \HofmanXT.
Note that this result is independent of the positions of the two
magnons on the transverse $S^3$, in accord with the expectation
of \HofmanXT.
The result \hmdt\ can also be read off directly from the solution
\hmsolution.

Another check is obtained by setting $\alpha_2 = \beta_2 = 0$
so that we have two dyonic giant magnons whose angular momenta both
lie within the $Z_2$ plane, leading to
\eqn\dtone{
\Delta T_{12} =
2 i { |1 - \lambda_1|^2 |1 + \lambda_1|^2 \over 
\lambda_1^2 - \lambdabar_1^2}
\log \left|
{\lambda_1 - \lambda_2 \over \lambda_1 - \lambdabar_2}
\right|
}
in complete agreement with \refs{\ChenGQ,\RoibanGS}
where this case was recently studied.

As a final consistency check, we can take $\alpha_2 = \beta_2 = \pi/2$,
which leads to
\eqn\dttwo{
\Delta T_{12} = 2 i { |1 - \lambda_1|^2 |1 + \lambda_1|^2 \over 
\lambda_1^2 - \lambdabar_1^2}
\log \left|
{ \lambda_1-1/ \lambdabar_2 \over  \lambda_1-1/ \lambda_2}
\right|.
}
{}From equation \onesoliton\ it is evident that
this choice of orientation simply reverses the sign of the
angular momentum of the second soliton
relative to $\alpha = \beta = 0$.  But this
is completely equivalent to changing $\lambda_2 \to 1/\lambdabar_2$, which
is indeed precisely the transformation between \dtone\ and \dttwo.

\bigskip
\noindent
{\bf Acknowledgements}

We are grateful to M. Abbott, A. Jevicki, and J. Maldacena
for useful discussions and suggestions.
The research of MS is supported by NSF grant PHY-0610529.
Any opinions, findings, and conclusions or recommendations
expressed in this material are those of the authors and do not
necessarily reflect the views of the National Science Foundation.

\listrefs
\end